\documentstyle [12pt,epsfig]{article}
\begin{document}
\begin{center}
\Large{\bf Effect of Finite Granularity of Detector on Anisotropy 
Coefficients}\\
\end{center}

\begin{center}

{\large{
\noindent
Sudhir~Raniwala$^1$, Marek~Idzik$^2$, Rashmi~Raniwala$^1$, Yogendra~Pathak~Viyogi$^{2,*}$\\
}}
\vskip 0.5cm
{\small{
$^1$~Physics Department, University of Rajasthan, Jaipur 302 004, India\\
$^2$~Faculty of Physics and Applied Computer Science,\\
AGH University of Science and Technology,
Krakow, Poland\\
$^3$~Variable Energy Cyclotron Centre,  Kolkata 700 064, India\\
\noindent $^{*}$ now at Institute of Physics, Bhubaneswar 751 005, India
}}

\end{center}

\abstract{
The coefficients that describe the anisotropy in the azimuthal distribution of 
particles are lower when the particles are recorded in a detector with finite 
granularity and measures only hits. This arises due to loss of information 
because of multiple hits in any channel. The magnitude of this loss of signal 
depends both on the occupancy and on the value of the coefficient. 
These correction factors are obtained for analysis 
methods differing in detail, and are found to be different.  
}%end of abstract

\begin{normalsize}
\section{Introduction}

Azimuthal anisotropy in particle emission in ultra-relativistic
heavy ion collisions was proposed as an important probe 
of the dynamics of the system ~\cite{ollie}. 
Subsequently, various methods have been proposed to obtain 
this anisotropy ~\cite{posk,olli-cumu,olli-leeyang}. 
In the more commonly used method the azimuthal distributions are 
expanded in a Fourier series where the coefficients of expansion 
are the measures of different orders of anisotropy ~\cite{posk}.
For small values of these coefficients, the first two terms describe 
an elliptic shape.
The first order anisotropy $v_1$, the directed flow, measures the shift of 
the centroid of the distribution and is the coefficient of the first 
term in the expansion. The second order anisotropy $v_2$, the elliptic flow,
measures the difference between the major and minor axes of the elliptic
shape of the azimuthal distribution and is the coefficient of the second 
term in the expansion.
The elliptic flow, $v_2$, probes the early stages of expansion of the 
interacting system and has been measured by large number of experiments
for different particle species in different 
kinematic domains for a variety of colliding systems and a range of 
center of mass energies ~\cite{ags,wa93,na49,wa98pmd,wa98leda,ceres,star,phenix,phobos}.
These measurements have provided new perspectives on the  
observed mass dependence of the elliptic flow ~\cite{sergei}. 
%The need for accurate determination of anisotropy 
%coefficients can not be overemphasised.

In detector sub systems where $p_T$ or energy is not measured, the 
anisotropy coefficients
are determined from the azimuthal distribution of the number of particles.
Some detectors measure the distribution of 
hits\footnote{One activated cell is counted as 
N$_{hits}$ = 1 irrespective of the number of tracks activating it.}.
If a cell is hit by more than one particle, information is lost because
the cell is still registered as one hit. 
The 
($\Delta \eta \times \Delta \Phi$) size of 
each cell in Silicon Pad Multiplicity Detector in WA98 experiment 
is about 0.07 $\times$ 2$^\circ$ ~\cite{spmd_nim}, whereas 
the corresponding size in the silicon strip detector in NA50 experiment 
is about 0.014 $\times$ 10$^\circ$ ~\cite{na50nim}.
For the same occupancy, both detectors will lose comparable number 
of particles by measuring hits. Therefore, the anisotropy coefficients 
describing distribution of hits are expected to 
be smaller than coefficients describing distribution of particles; 
$v_n^{hits} < v_n$  and there is need to  determine 
an appropriate correction factor which will depend on the 
granularity of the detector. 
Methods to estimate the effect of finite granularity
have been discussed in ~\cite{aliceint}. 
In the present work
the effect of finite granularity on the standard methods of analysis is
investigated by folding the detector geometry in the simulated data.

In Section 2, an approximate expression for the ratio 
$\frac {v_n^{hits}}{v_n}$ is obtained as a function of occupancy. 
Section 3 describes the simulation and the different methods of analysis. 
The Results are discussed in Section 4.

\section{Multiple Hits and Azimuthal Anisotropy}

It is possible to have an ideal setup for a simulation
experiment for any conceivable granularity. However, an actual experiment 
has a finite number of detector cells $N_{cells}$, which defines 
the coarseness of the
granularity for a given acceptance of the detector. For a given average 
number of 
incident particles, $<N_{part}>$, one can define the mean 
occupancy $\mu_0$ 
\begin{eqnarray}
	\mu_0 = \frac{<N_{part}>}{N_{cell}}
\label{occu}
\end{eqnarray}

Using the Poisson distribution for probability of $n$ particles 
incident on any cell, one can deduce the average number of hits as 
%\begin{eqnarray}
%	P_n  = \frac{\mu_0^n \cdot e^{-\mu_0}}{n!}
%\end{eqnarray}
% This gives 
\begin{eqnarray}
	\frac{<N_{hits}>}{<N_{part}>}  =  
%\frac{\Sigma P_n }{\Sigma n \cdot P_n }
			      \frac{1 - e^{-\mu_0}}{\mu_0} = \frac{-x}{ln(1-x)}
\label{hits}
\end{eqnarray}
 where $x$ = $<N_{hits}>/N_{cell}$ is the hit-occupancy and is 
experimentally measurable. 
 Mean occupancy can also be written as $\mu_0 = - ln(1-x)$.
Since the total number of cells ($N_{cell}$) is the sum of the 
average 
number of occupied ($<N_{occ}>$)
and unoccupied ($<N_{unocc}>$) cells,
one can immediately obtain the expression for occupancy as in 
ref~\cite{phobos}.
\begin{eqnarray}
	\mu_0 = ln ( 1 + \frac{<N_{occ}>}{<N_{unocc}>})
\end{eqnarray}
enabling its determination from experimentally measurable quantities.

%\section{Effect on Anisotropy Coefficients}

The anisotropy in the azimuthal distribution of the number of incident  
particles is written as 
\begin{eqnarray}
 N_{part}(\phi) = <N_{part}>( 1+ \sum 2v_n \cos n (\phi - \psi_n) )
\label{azim}
\end{eqnarray}
where $\psi_n$ is the event plane angle. Using equations 
~\ref{occu} and ~\ref{hits}, one can write the azimuthal dependence
of hits as 
\begin{eqnarray}
	N_{hits}(\phi)  \propto 1-e^{-\mu(\phi)}
\label{phihit}
\end{eqnarray}
\noindent where $\mu(\phi)$ denotes the azimuthal dependence of the occupancy.
Since the intrinsic occupancy of cells increases with the 
increase in 
the number of incident particles, the occupancy will have the same azimuthal
dependence as the incident particles and can be written as
$\mu(\phi)$ = $\mu_0 (1+ \sum 2v_n \cos n (\phi - \psi_n) )$.
%Substituting 
%equation ~\ref{azim} in equation ~\ref{hits}  
%gives the azimuthal dependence of the hits as 
%\begin{eqnarray}
%	N_{hits}  \propto 1-e^{-(1+ \sum 2v_n \cos n (\phi - \psi_n) )} 
%\end{eqnarray}
Substituting this in equation ~\ref{phihit} and expressing $N_{hits}$ as 
a Fourier series with coefficients $v_n^{hits}$  
enables a determination
of the ratio $v_n^{hits}/v_n$. To the first order, this ratio can be 
approximated as 
\begin{eqnarray}
  \frac {v_n^{hits}}{v_n} = \frac {1 - \mu_0 + \frac {\mu_0^2}{2} - f(v)}{1 - \frac {\mu_0}{2} + \frac{\mu_0^2}{6}} 
\label{corfac1}
\end{eqnarray}
where $f(v)  = \mu_0v_2$ for $n = 1$, and = $ \frac {\mu_0v_1^2}{2v_2}$ for $n = 2$.

%\begin{itemize}
%\item $f(v) = \mu_0v_2$ for $n = 1$.
%\item $f(v) = \frac {\mu_0v_1^2}{2v_2}$ for $n = 2$.
%\end{itemize}

The function $f(v)$ contributes little for small values of 
occupancy and flow.
%\footnote{In nature, flow decreases with increasing centrality} 
%and is $\mu_0v_2$ for $n = 1$ and is $\frac {\mu_0v_1^2}{2v_2}$ for 
%$n = 2$.

The ratio can also be approximated as  
\begin{eqnarray}
  \frac {v_n^{hits}}{v_n} = - \frac {1-x}{x} \cdot ln(1-x) 
\label{corfac2}
\end{eqnarray}
%where $x$ =  $\frac{N_{hits}}{N_{cell}}$ (=$\frac{N_{occ}}{N_{cell}}$).

These results have been applied to the data recorded in the Silicon 
Pad Multiplicity Detector in the WA98 experiment ~\cite{wa98pmd}.
The results from equation ~\ref{corfac1} and ~\ref{corfac2}
are corroborated with results from 
simulations as described in the following.

\section{Simulation and Analysis}
         
%Simulation experiments involve two steps: event generation and 
%analysis. For the purpose of the present work, one needs
%to know the azimuthal distribution of the hits and of the particles
%for each event, since the analysis is based only on these variables.
%\subsection{Event Generation}          
          
%Multiplicities for the events are generated so that these are 
%commensurate with the range of occupancies in various experiments.
For the present simulation experiments, the events have been generated with
various values of charged particle multiplicity corresponding to different 
occupancies in the detector. Assuming a constant dN/d$\eta$ and an 
exponential p$_T$ distribution, the kinematic variables of 
each particle are generated with p$_T$ in the region 
0 to 6 GeV/c and  $\eta$ in an assumed region of acceptance of the detector.
Typical ranges of $\eta$ chosen in the present work vary between 0.5 and 1.0.
Azimuthal angle of each particle is assigned according to the probability 
distribution ~\cite{rashmi}
\begin{eqnarray}
      r(\phi)= \frac{1}{2\pi}\left[1 + 2v_1 \cos(\phi -\psi_R) +
 2v_2 \cos 2(\phi-\psi_R)\right]
\label{aniso}
\end{eqnarray}

\noindent where $\psi_R$ is randomly generated once for each event.
Events are generated for different granularities in 
$\eta$ and $\Phi$. A constant dN/d$\eta$ distribution
and cells of equal $\Delta \eta$ intervals give a uniform 
intrinsic occupation  
probability for each cell.
Detector geometry, flow and occupancy
are varied for a systematic study.

In the present work, the granularities that are chosen are 
fairly arbitrary but commensurate with the coarseness of certain detectors
~\cite{spmd_nim,na50nim}. More specifically, simulations are performed for 
%\begin{itemiz}
$\Delta \eta$ = 0.07, $\Delta \Phi$ = 2$^\circ$;
$\Delta \eta$ = 0.014, $\Delta \Phi$ = 10$^\circ$; 
$\Delta \eta$ = 0.00875, $\Delta \Phi$ = 10$^\circ$. 
%\item $\Delta \eta$ = 0.004375, $\Delta \phi$ = 10$^\circ$ (4608 channels)
%\end{itemize}
The results are based on an analysis of 10$^6$ events in each case.
For low flow values ($v_n$ = 0.02), the number of generated events 
is 2.5 $\cdot$ 10$^6$.

\subsection{Different Data-Sets}
         
The number of particles simulated and the number of 
cells activated ($N_{hits}$) are known for each event.
It is assumed that one incident particle does not activate more 
than one cell.
         
The following information is stored as three different data-sets from 
the simulated events.
%\begin{enumerate}
\begin{description}
\item[(a)] the number of particles and the azimuthal angle of each 
particle. This corresponds to a measurement in an ideal detector of
infinite granularity. Anisotropy coefficients measured thus are labeled as $v_n^{ideal}$.

\item[(b)] the number of particles and the azimuthal angle of each hit cell. 
This corresponds to the case when azimuthal angle of each particle is known to 
an accuracy determined by the azimuthal size of the cell and the number of 
particles can be determined using the pulse height information. 
This is equivalent to randomly adding (or subtracting)
$\delta \phi$ ( $\leq \frac {\Delta \Phi}{2}$) to each $\phi$ where 
$\Delta \Phi$ is 
the azimuthal size of each cell.
%There is a random indeterminacy in each particle 
%angle up to a maximum of $\pm \Delta \Phi/2$. 
The anisotropy coefficients obtained using these are called $v_n^{ch}$ and 
can be written as ~\cite{starresult} 
\begin{eqnarray}
 \frac {v_n^{ch}}{v_n^{ideal}} = \frac {\sin n\frac {\Delta \Phi}{2}}{n\frac{\Delta \Phi}{2}}
\label{starres}
\end{eqnarray}
\item[(c)] the number of hits and azimuthal angle of each hit cell. This
is the information recorded by the detectors that produce only a binary
(hit/no-hit) signal for each cell and the corresponding anisotropy coefficients
are called $v_n^{hits}$. 
\end{description}

\subsection{Methods of Analysis}

Fourier coefficients  of $n^{th}$ order can be determined from the 
azimuthal distribution of the particles with respect to the event 
plane angle of order $m$, provided $n$ is an integral multiple 
of $m$, by fitting to the following equations ~\cite{posk}
\begin{eqnarray}
\frac {dN}{d(\phi-\psi'_m)}& \propto  1 + \sum_{n=1}^{\infty} 2v'_{nm} \cos nm (\phi -\psi'_m)  
\label{firstord}
\end{eqnarray}
The event plane angle is given by 
\begin{eqnarray}
      \psi'_m= \frac{1}{m}\left(\tan^{-1} \frac{\Sigma w_i \sin m\phi_i}
        {\Sigma w_i \cos m\phi_i}\right)
\label{event}
\end{eqnarray}
\noindent where the summation is over all particles $i$ and the 
weights $w_i$ are all set to 1. 
The average deviation of the estimated event plane 
from the true event plane due to multiplicity fluctuations can be  
determined experimentally and is termed as the resolution correction 
factor (RCF). 
Experimentally, RCF is obtained using 
the  sub-event method  described in reference ~\cite{posk}. 
Here every event is  divided into two 
sub-events  of equal multiplicity and the event plane angle $\psi'_m$ 
is determined for each sub-event.
This enables determination 
of a parameter $\chi_m$ directly from the experimental data
using the fraction of events where the correlation of the planes 
of the sub-events is greater than $\pi/2$ ~\cite{posk,ollieconf}:
\begin{eqnarray}
        \frac{N_{events} ( m | {\psi_m'^a - \psi_m'^b|} > \pi/2)}{N_{total}} 
= \frac{e^{-\frac{\chi_m^2}{4}}}{2}
\label{chim}
\end{eqnarray}
where $N_{total}$  denotes the total number of
events, $\psi_m'^a$, $\psi_m'^b$ are the observed  
event plane angles of the two sub-events (labeled
$a$ and $b$) and the numerator on the left denotes the
number of events having the angle between sub-events greater
than $\pi/2m$. 
The parameter $\chi_m$ is used to determine 
RCF$_{nm}$ = $\langle \cos(nm(\psi'_m-\psi^{true}_m))\rangle $,
where $\psi^{true}_m$ is the true direction of the 
event plane, and the average is over all events. 
The RCF can be determined from  $\chi_m$ by the following 
relation in reference ~\cite{posk}.

\begin{eqnarray}
 \nonumber \langle \cos(nm(\psi'_m-\psi_m^{true}))\rangle = \frac{\sqrt{\pi}}{2\sqrt{2}} {\chi^{}_{m}} \exp(-\chi^2_m/4) \cdot \\
%\left[{\em{I }}_{\frac{m-1}{2}}(\chi^2_n/4) + {\em{I}}_{\frac{m+1}{2}} ( \chi^2_n/4 ) \right]
\left[{{I }}_{\frac{n-1}{2}}(\chi^2_m/4) + {{I}}_{\frac{n+1}{2}} ( \chi^2_m/4 ) \right]
\label{rcfeqn}
\end{eqnarray}

\noindent where $I_\nu$ are the modified Bessel functions of order $\nu$. The 
RCF can also be obtained by obtaining 
$\langle \cos n (\psi'^a_n - \psi'^b_n) \rangle$, where $\psi'^{a,b}_n$ are the 
event plane angles of the two sub-events.

%\subsection{Anisotropy Coefficients}

In the present work, Fourier coefficients $v'_{nm}$ are
extracted for the case with event plane
order equal to the order of the extracted Fourier 
coefficient, i.e. $v'_{nm}$ = $v'_{nn}$ and is denoted here by $v'_n$.
The values of $v_n$ have been obtained by the following methods: 
\begin{description}
\item[\bf{Method 1 : }] 
In this method, the sub-events for each event 
were formed by dividing the pseudorapidity range into two such that 
each sub-event has equal number of particles(hits)\footnote{The two 
sub-events were also formed by assigning 
particles(hits) to each from alternate segments of the azimuthally segmented 
detector.}. Then
$v'^a_n$ = $\langle \cos n (\phi_i^a - \psi'^b_n) \rangle $  and
$v'^b_n$ = $\langle \cos n (\phi_i^b - \psi'^a_n) \rangle $ are determined
where $\phi_i^a$ represent the azimuthal angles of particles in sub-event $a$ and 
$\psi'^b_n$ is the event plane angle determined using particles in
sub-event $b$. The averages
are computed over all particles over all events. In the absence of 
non flow correlations
\begin{eqnarray} 
v_n = \sqrt \frac {v'^a_n \cdot v'^b_n}{\langle \cos n (\psi'^a_n - \psi'^b_n) \rangle}
\label{geommean}
\end{eqnarray} 

It is also possible to obtain $v'_n$ by fitting equation ~\ref{firstord} 
to the $\phi_i - \psi'_n$ distribution. This distribution is a sum of the 
the distributions $\phi^a_i - \psi'^b_n$ and $\phi^b_i - \psi'^a_n$. 
This yields 
\begin{eqnarray}
        v_{n} = \frac{v'_{n}}{\sqrt{\langle \cos(n(\psi'^a_n-\psi'^b_n))\rangle}} 
\label{corrfact}
\end{eqnarray}
The denominator in both cases above is the event plane resolution 
correction factor when the 
event plane is determined for the sub-event with half of the complete
event multiplicity and is approximately lower than the full event RCF
by a factor $\sqrt{2}$.  
The values of $v_n$ obtained this way are termed as $v_n^{geom}$.

\item[\bf{Method 2 : }] 
In this method, in each event, the sub-events were formed by randomly 
selecting one half of the particles. 
The $v'_n$  values are also extracted by fitting equation ~\ref{firstord}
to the $\phi - \psi'_n$ distribution, where $\psi'_n$ is obtained by 
excluding the 
particle (or hit) being entered in the 
distribution\footnote{This avoids autocorrelations but also introduces 
a negative 
correlation, effectively decreasing the values of $v'_n$.}.
The $v_n$ values are obtained from
\begin{eqnarray}
        v_{n} = \frac{v'_n}{RCF_{n}} 
\label{corrfact2}
\end{eqnarray}
$RCF_n$ is the resolution correction factor for the full event plane
and is obtained by the correlation between randomly divided 
sub-events of equal multiplicity and using equations ~\ref{chim} and 
~\ref{rcfeqn}.
The values of $v_n$ obtained this way are termed as $v_n^{rand}$.
\end{description}

\section{Results and Discussion }

A relation for the ratio of the 
anisotropy measured using hits to the actual anisotropy 
for different values of occupancy was obtained in section 2.
These values are
corroborated using simulations and the results are discussed in 
this section. The different methods of analysing the 
data discussed in section 3 are applied 
on simulated data to study the effect of finite
granularity on the $v_n$ values.
The simulated data are analysed for all the detector geometries 
described above. The results are discussed for the case
$\Delta \eta$ = 0.07, $\Delta \Phi$ = 2$^\circ$. The conclusions 
remain the same for the other geometries. 
For all simulations, N$_{hits}$/N$_{part}$ is obtained for 
different values of $x$ and is found to be consistent with
results from equation ~\ref{hits}.

\subsection{{\bf $v_n$} Using Known Event Plane: Actual Dilution} 

The dilution in the anisotropy coefficients due to finite granularity
can be computed using the known event plane angle ($\psi_n$) in simulation.
The quantity  
$v_n$  = $\langle \cos(n(\phi_i-\psi_n))\rangle $ is determined for the
different data-sets described in Section 3.1 and yields the finite 
granularity effect on the anisotropies in the distribution. 

%\begin{enumerate}
%\item $v^{ideal}_n$ using all particles and their 
%angles. This corresponds to measurements using a detector with 
%infinite granularity.
%\item  $v^{ch}_n$ using all particles but their angles are 
%replaced by the angles corresponding to the mean positions
%of the detector cells. This is equivalent to randomly adding (or subtracting)
%$\delta \phi$ ( $\leq \frac {\Delta \Phi}{2}$) to each $\phi$ where 
%$\Delta \Phi$ is 
%the azimuthal size of each cell.
%\item $v^{hits}_n$ using all hits and the angles corresponding to 
%the mean positions of the detector cells.
%\end{enumerate}

\begin{enumerate}         
\item  $v^{ideal}_n$ reproduces the input flow, as expected naively.

\item The dilution due to coarse information about the particle angle 
can be judged by plotting the ratio $v^{ch}_n/v^{ideal}_n$. 
The result for the granularity 
$\Delta \eta \times \Delta \Phi = 0.07 \times 2^\circ$
has been plotted in Fig.~\ref{vtheory} for two different values of 
initial anisotropy. 
The correction factor due to coarse
information of particle angle is very close to 1 for such a  
small azimuthal size of the cells. 
The results have been corroborated using simulations for values of 
$\Delta \Phi$ up to $30^\circ$, and agree with results from equation 
~\ref{starres}.

\item The anisotropy for the hits, $v^{hits}_n$ is diluted both due to 
coarse information of particle angle and loss of particles because of 
the multiple hits.  
The resultant loss is 
best seen by 
plotting the ratio $v^{hits}_n/v^{ideal}_n$ as a function of hit-occupancy 
for the simulated data. 
The results are shown in Fig.~\ref{vtheory} for a 
$\Delta \Phi = 2^\circ$ along with the estimates obtained using equations
~\ref{corfac1} and ~\ref{corfac2}. 
For $\Delta \Phi = 2^\circ$, results in (ii) above show that the coarse 
information of particle angle has very little effect, and the dilution 
in $v^{hits}_n$ is primarily due 
to loss of particles because of multiple hits. 
Simulation results corroborate the 
analytical expression that include a weak dependence on anisotropy $v$.
\end{enumerate}

\begin{figure*}[h] 
\centerline{\includegraphics{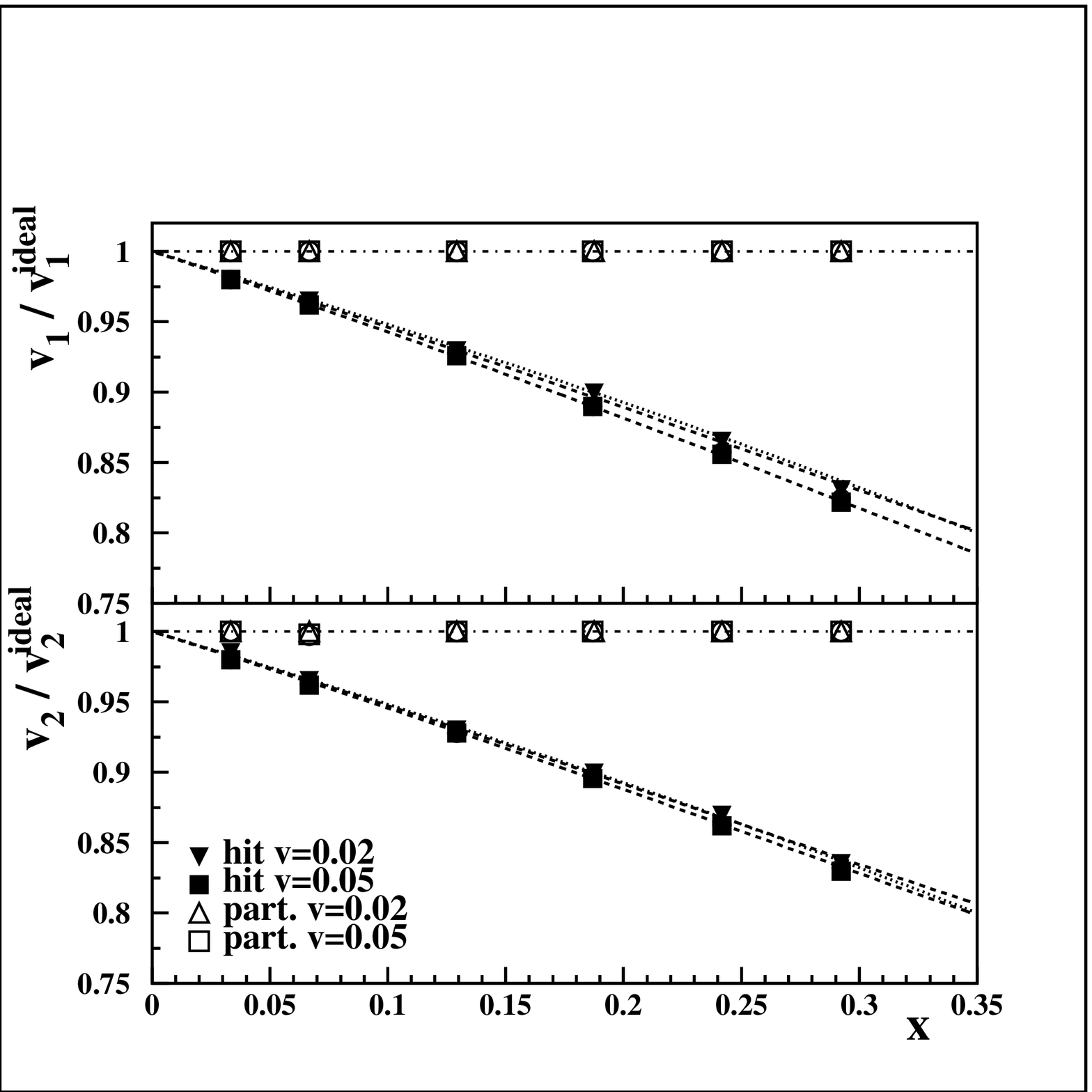}}
\caption{ Open symbols show the ratio $v^{ch}_n/v^{ideal}_n$ 
for different values of $x$, where $x$ = $<N_{hits}>/N_{cell}$ 
is experimentally measurable. 
Filled symbols show the ratio 
$v^{hits}_n/v^{ideal}_n$. Squares are for $v_n$ = 0.05 and 
triangles are for $v_n$ = 0.02. 
The top panel is 
for $v_1$ and the bottom panel is for $v_2$.
The two dashed curves in each panel correspond to the different values of 
anisotropy $v_n$ and represent equation ~\ref{corfac1}.
The dotted curve represents equation ~\ref{corfac2}.
A horizontal line at the value of ratio equal to 1 is also drawn.}
\protect\label{vtheory}
\end{figure*}

\subsection{{\bf $v_n$} Using Reconstructed Event Plane: Observed Dilution}

In this section we investigate the results on dilution of anisotropy when 
the event plane and its resolution is determined from the data.
Both the methods listed in Section 3 require a determination of 
(i) the uncorrected $v'_n$ and (ii) the corresponding (sub) event plane resolution.
The quantitative effect of finite granularity 
on each of these quantities is different, and hence the measured 
values of $v_n$ are different from the initial values.
The results of a systematic investigation 
are shown in figure ~\ref{rcfetc} for varying 
hit-occupancies.  The results from method 1 of section 3 are shown in the left
column and those from method 2 are shown in the right column. For both 
methods, the results from data-sets (b) and (c) are scaled by the 
corresponding values obtained using data-set (a). 
The open circles show the results obtained using the data-set (b) for 
all charged particles and the mean angles of the cell positions. 
The filled circles show the results obtained using the data-set (c)
for the hits and the corresponding angles. 
For both methods, analysis of data-set (a) reproduces the initial
anisotropy, validating the methodology.

\subsubsection{Division into sub-events based on geometry: {\bf $v_n^{geom}$}}

The anisotropy coefficients $v_n^{geom}$ determined using 
equations ~\ref{geommean} and 
~\ref{corrfact} yield identical results. 

The event plane resolution correction factor and the uncorrected values
of $v'_1$ are seen to decrease by different factors due to finite 
granularity effect, resulting in a reduced value of $v_1^{geom}$.  

%Equations
%~\ref{corfac1} and ~\ref{corfac2} are shown as solid and dotted 
%line respectively. The dashed line represents equation ~\ref{corfac1} without
%the term $f(v)$. One observes that simulation
%results and the corrections factors obtained above corroborate each other.
The results show that the value of anisotropy 
measured using a detector with 30\% mean hit-occupancy is to be corrected 
by a factor of $\sim$ 1.2 to obtain the actual value of
anisotropy ($v_1$), clearly a significant effect. The results for the second 
order anisotropy are similar, with small quantitative difference as seen
from equation ~\ref{corfac1}.

\begin{figure*}[h] 
\centerline{\includegraphics{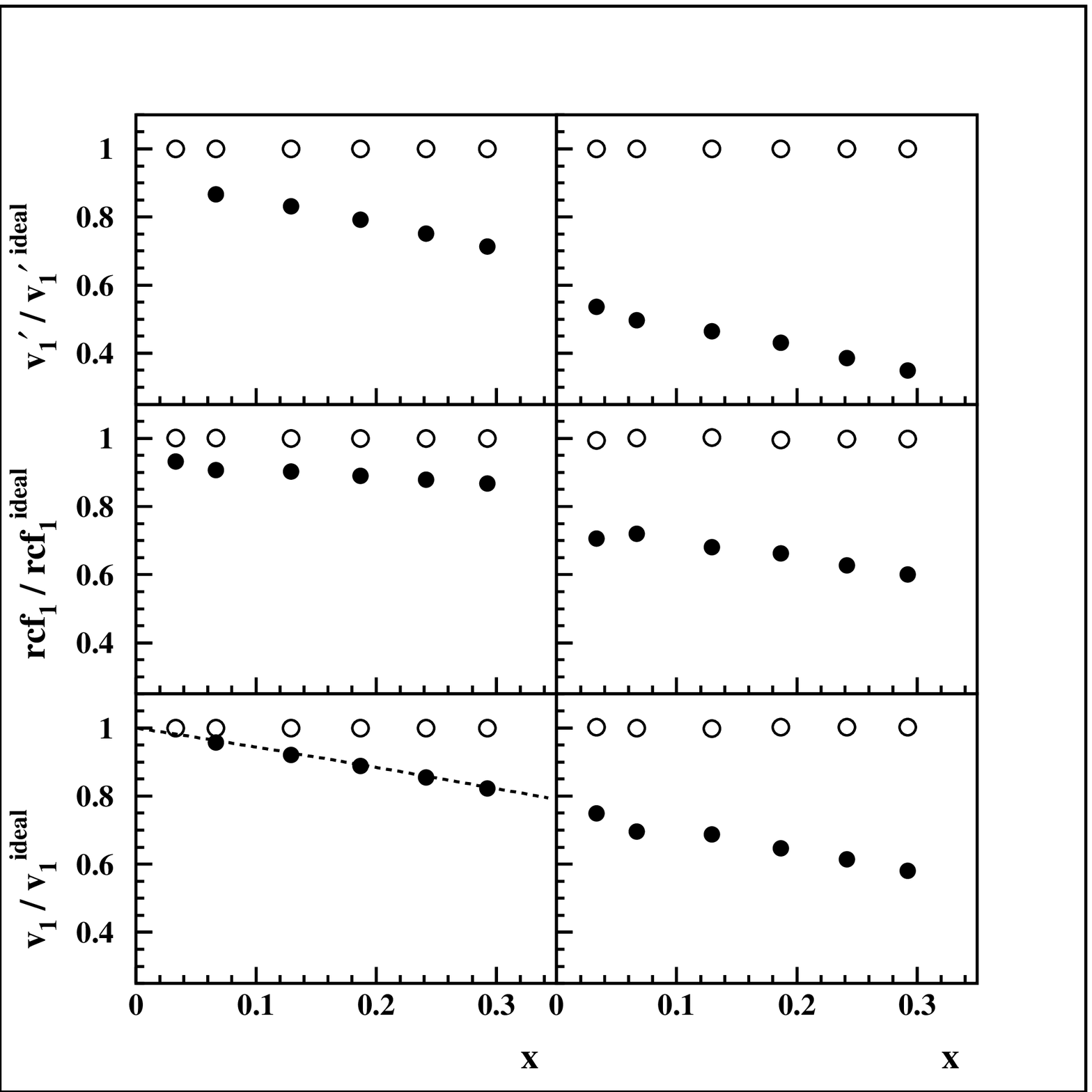}}
\caption{ 
The three panels in the left column show uncorrected values of $v'_1$, $RCF_1$ and 
$v_1^{geom}$ for charged particles and hits for an initial $v_1$ = 0.04
for different values of hit-occupancy $x$.
The open 
circles are for all particles and the filled circles are for hits,
as described in the text. The values
are scaled with corresponding values for the ideal case.
The dashed curve represents equation ~\ref{corfac1}. 
The panels in the right column show the corresponding results for method 2, 
the random subdivision of events with the bottom column showing the value of 
$v_1^{rand}$ scaled by the corresponding value for the ideal case.   
}
\protect\label{rcfetc}
\end{figure*}

The analysis described as Method 1 was repeated for the case when the two 
sub-events were formed
by assigning particles(hits) to each from alternate segments of the azimuthally
segmented detector. The results remain the same. 

\subsubsection{Random division into sub-events: {\bf $v_n^{rand}$}}

The right column of Figure ~\ref{rcfetc} shows the results for method 2 
when the 
events are divided randomly into two sub-events, and the projection of 
particles/hits is taken on the event plane angle of the full event
(after removing autocorrelations). 
This method works for the simulated data corresponding to data-sets (a) 
and (b) described in Section 3.1, and the values of data-set (b) scaled 
to the corresponding values from data-set (a) are shown as open symbols in
the right column of Fig. ~\ref{rcfetc}.  The values obtained 
using data-set (c) are much lower than the values obtained by analysing
the same data using Method 1. The systematically lower
values of the quantities for Method 2 arise due to 
multiple hits, due to removing auto correlations and 
due to random division into sub-events. The combined 
effect results in much lower values of $v_1^{rand}$.
For data-set (b), the method of removing autocorrelation removes only 
one particle while the other neighbouring particles are used in 
determining the event plane. For data-set
(c) comprising of hits, one detector cell is removed from the data that 
determines the event plane, effectively removing all particles within 
the azimuthal size of that cell, introducing a negative correlation,
resulting in much lower values of $v'_1$.
This holds true for all values of occupancy. 

The decrease in the values of event plane resolution can be understood 
as follows:
For an azimuthal distribution given by equation ~\ref{aniso},
the particle density is maximum 
along the direction of the reaction plane. On an event by event 
basis, the maximum loss of particles due to multiple hits 
will be along this direction. Consider there are $N_{corr}$ correlated 
particles
in a region $\delta \phi$ about the reaction plane, where $N_{corr}/N_{total}$
is greater than $\delta \phi/2 \pi$. When such an event is 
divided into two equal multiplicity sub-events, the correlation between the 
two sub-events will be maximum if $N_{corr}$/2 particles go into
each sub-event.
Though this is true on the average, {\it on an event 
by event basis}, only a certain number out of $N_{corr}$ fall 
into one sub-event. The correlation between sub-events for these events
is less than the corresponding situation described above. 
The correlation will be weakest if all of 
these particles fall into one sub event. 
In such a situation,
the $v_n^{rand}$ values are likely to be much lower than the $v^{ideal}_n$.
However, the probability of the random division into sub-events leading 
to this situation is $(1/2)^{N_{corr}-1}$ and is small.

The situation remains the same when hits are recorded instead 
of particles, and $N_{corr}$ is replaced by $N_{corr(hits)}$,
and $N_{total}$ by $N_{hits}$. The probability of 
a random division with all $N_{hits}$ going into a sub event
is  $(1/2)^{N_{corr(hits)}-1}$. This probability is clearly
greater than the corresponding case where there is no loss of particles
due to multiple hits. 
When this happens, the two sub-events show little correlation,
causing both $v'_n$ and RCF$_n$ to decrease. The decrease in both
quantities only partially compensate each other and the resultant values 
are observed to be lower, as shown in the third panel on the right column
of figure ~\ref{rcfetc}.

The resultant effect is to produce a reduced value of $v_n^{rand}$,
which would need a relatively larger correction factor to obtain
the original value. 
Repeating the simulation for different values of anisotropy 
shows that the correction factor increases with decreasing values of 
anisotropy. Below a certain value of anisotropy, $v^{thres}$, the values 
of anisotropy extracted using this method yields $v_1^{rand}$ consistent 
with zero, putting a limit on the sensitivity of detecting 
anisotropy in the data. The values of $v^{thres}$ depend both on 
the granularity of the detector and on the order of the anisotropy being 
determined.

\section{Conclusions}

The anisotropy in the distribution of hits is shown to be 
lower than the anisotropy
in the distribution of particles.
Loss of particle information
due to multiple hits (or two track resolution) contributes significantly 
to the dilution of the observed anisotropy values. The correction factor
for the dilution has been obtained and confirmed by simulation 
experiments for different detector geometries. Different methods
of event subdivision yield same results for an ideal detector.
While it is known that the event subdivision obtained randomly 
does not work in the presence of non-flow correlations, the limitation
of this method is shown here for the case where there are no
non-flow correlations. The anisotropy values $v_n^{rand}$ obtained using 
the distribution of hits are much lower than the corresponding values of 
$v_n^{geom}$ and need 
a correction factor which is larger than the one obtained in 
Equation ~\ref{corfac1}.
 The analysis and the discussion in the present work are suited for 
the case where the event plane is being determined from the same 
set of particles. 
The event plane resolution correction factor corrects only
for fluctuations arising due to finite multiplicity.
If the anisotropy parameters are determined with respect to an event plane
determined from another set of particles measured using a detector
with different granularity, then the correction factors need to be 
determined differently and their determination is outside the scope 
of the present work.
% (This is necessary because the 
%the event plane resolution correction factor corrects only for 
%fluctuations in multiplicity, and not for coarse granularity).

\vskip 0.5cm
\end{normalsize}

\noindent 
{\large{\bf Acknowledgement}}

\begin{sloppypar}
The financial support from the Department of Science and Technology and the 
Department of Atomic Energy of the Government of India is gratefully
acknowledged.

\end{sloppypar}

\end{document}